# Illumination and annealing characteristics of two-dimensional electron gas systems in metal-organic vapor-phase epitaxy grown Al$_x$Ga$_{1-x}$N/AlN/GaN heterostructures

N. Biyikli,[a] Ü. Özgür, X. F. Ni, Y. Fu, and H. Morkoç

*Department of Electrical & Computer Engineering, Virginia Commonwealth University, Richmond, VA 23284*

Ç. Kurdak

*Department of Physics, University of Michigan, Ann Arbor, MI 48109*

## ABSTRACT

We studied the persistent photoconductivity (PPC) effect in Al$_x$Ga$_{1-x}$N/AlN/GaN heterostructures with two different Al-compositions (x=0.15 and x=0.25). The two-dimensional electron gas formed at the AlN/GaN heterointerface was characterized by Shubnikov-de Haas and Hall measurements. Using optical illumination, we were able to increase the carrier density of the Al$_{0.15}$Ga$_{0.85}$N/AlN/GaN sample from $1.6 \times 10^{12}$ cm$^{-2}$ to $5.9 \times 10^{12}$ cm$^{-2}$, while the electron mobility was enhanced from 9540 cm$^2$/Vs to 21400 cm$^2$/Vs at T = 1.6 K. The persistent photocurrent in both samples exhibited a strong dependence on illumination wavelength, being highest close to the bandgap and decreasing at longer wavelengths. The PPC effect became fairly weak for illumination wavelengths longer than ~530 nm and showed a more complex response with an initial negative photoconductivity in the infrared region of the spectrum ($\lambda$ > 700 nm). The maximum PPC-efficiency for 390 nm illumination was 0.011% and 0.005% for

[a] Electronic mail: nbiyikli@vcu.edu

$Al_{0.25}Ga_{0.75}N/AlN/GaN$ and $Al_{0.15}Ga_{0.85}N/AlN/GaN$ samples, respectively. After illumination, the carrier density could be reduced by annealing the sample. Annealing characteristics of the PPC effect were studied in the 20 – 280 K temperature range. We found that annealing at 280 K was not sufficient for full recovery of the carrier density. In fact, the PPC effect occurs in these samples even at room temperature. Comparing the measurement results of two samples, the $Al_{0.25}Ga_{0.75}N/AlN/GaN$ sample had a larger response to illumination and displayed a smaller recovery with thermal annealing. This result suggests that the energy scales of the defect configuration-coordinate diagrams for these samples are different, depending on their Al-composition.

## I. INTRODUCTION

Because of their potential applications in high-temperature, high-power, and high-frequency microwave electronics, $Al_xGa_{1-x}N/GaN$ heterostructures are attracting escalating research interest [1–5]. Promising device performances have been reported as well as continuously improving carrier transport properties. Better quality substrates, improved growth techniques and original layer structures are being used to realize the full potential of this material system [6–10]. Understanding and minimizing defects are essential for further progress. The electronic defect state of the $Al_xGa_{1-x}N/GaN$ heterostructures can be altered by illumination of the sample at low temperatures. Associated with the change in defect configuration is the persistent photoconductivity (PPC) effect. Typically this effect leads to conductivity enhancement under optical illumination, which persists for a long period of time after the termination of optical excitation. PPC may have an adverse effect on stable device operation; however, it also provides a powerful tool to study the effect of carrier density on the transport properties.

PPC has been observed in most III-V semiconductor alloys including the $Al_xGa_{1-x}As$ material system, in which the effect was attributed to the excitation of the donor-vacancy (DX) centers [11–13]. The PPC effect observed in GaN and $Al_xGa_{1-x}N$ epilayers was similarly attributed to defect complexes such as gallium vacancies, nitrogen antisites, deep-level impurities, and interacting defect complexes [14-23]. For $Al_xGa_{1-x}N/GaN$ heterostructures, several groups have reported PPC experiments [24-26]. The persistent increase in carrier concentration was explained by the transfer of photo-excited electrons from deep-level impurities in $Al_xGa_{1-x}N$ layers [24]. However, the spectral illumination dependence of PPC in $Al_xGa_{1-x}N$/GaN heterostructures has not been investigated. In addition, a systematic study of the thermal annealing effect, which returns the sample back to its original low-carrier-density state, would be of significant interest. Our motivation in this work is to explore the spectral illumination and thermal annealing characteristics of the PPC effect in $Al_xGa_{1-x}N$/AlN/GaN heterostructures. We present our experimental results on the influence of PPC in high-mobility $Al_xGa_{1-x}N$/AlN/GaN heterostructures grown by metal-organic vapor-phase epitaxy (MOVPE). Using the PPC effect, we extracted the relation between the carrier density and mobility for the heterostructures under investigation. The spectral illumination and thermal annealing characteristics of the PPC effect and its variation with Al-composition were also studied systematically.

## II. EXPERIMENTAL DETAILS

The $Al_xGa_{1-x}N$/AlN/GaN heterostructures were grown by custom low-pressure metal-organic vapor-phase epitaxy (LP-MOVPE) in a rotating-disk vertical MOVPE chamber. Ammonia ($NH_3$), trimethylgallium (TMGa), and trimethylaluminum (TMAl) were used as precursors for nitrogen, gallium, and aluminum, respectively. Hydrogen was used as the carrier gas and growth was accomplished under high-speed rotation (~500 rpm). The samples were grown on c-plane (0001) sapphire substrates. The epitaxial layer structure of the samples consisted of a thick (~3 μm) GaN layer, followed by a ~1 nm thick AlN interfacial layer, a ~25 nm $Al_xGa_{1-x}N$ layer, and a ~3 nm GaN cap layer, all nominally undoped. We used an AlN interfacial layer to reduce the alloy disorder scattering by minimizing the wavefunction penetration from the 2DEG channel into the

$Al_xGa_{1-x}N$ layer [27, 28]. To study the effect of Al-composition in the $Al_xGa_{1-x}N$ layer, two samples were grown with x=0.25 (sample I) and x=0.15 (sample II). The growth was initiated with a 25-nm thick low-temperature (~550 °C) GaN nucleation layer. The 3 μm thick GaN epilayer was grown at 1010 °C under 200 mTorr and using a V/III ratio of ~4000. The chamber pressure and ammonia flow rate were decreased to 30 mTorr and 2 liter/min respectively, for the growth of $AlN/Al_xGa_{1-x}N$ layers. The growth temperature for these layers and the GaN cap layer was ~1060 °C.

Standard photolithography and reactive ion etching were used to define 600 μm long and 100 μm wide Hall-bar structures. Ohmic contacts were formed using Ti/Al/Ti/Au (30 nm/100 nm/30 nm/50 nm) alloyed at 900 °C for 1 min.

## III. RESULTS AND DISCUSSION

## A. Magneto-transport measurements

To confirm the presence of a 2DEG and to extract the carrier density and mobility of the samples, magneto-transport measurements were performed using a variable temperature liquid-He cryostat equipped with a superconducting magnet. Shubnikov-de Haas (SdH) and Hall measurements were carried out at T = 1.6 K within a magnetic field range of 0–6.6 T. Figure 1(a) shows the longitudinal resistivity ($\rho_{xx}$) and transverse (Hall) resistivity ($\rho_{xy}$) data measured at 1.6 K for sample I. Well-resolved magnetoresistance oscillations commencing around 2 T were observed, confirming the existence of a high-quality 2DEG. Quantum Hall steps in the Hall-resistivity curve accompany the SdH oscillations. The SdH oscillations without any beat characteristics indicated that only a single subband was occupied. Using $\rho_{xy}$ and $\rho_{xx}$ the 2DEG carrier density and mobility were determined to be $n=3.0\times10^{12}$ cm$^{-2}$ and $\mu=11710$ cm$^2$/Vs at 1.6 K, respectively. Magneto-transport measurements of sample II revealed a lower carrier density and a lower mobility: $n=1.6\times10^{12}$ cm$^{-2}$ and $\mu=9540$ cm$^2$/Vs at 1.6 K. The SdH data of both samples are compared in Figure 1(b). Sample II exhibited lower carrier density and higher sample resistance, which is a result of lower Al-composition in the Al$_x$Ga$_{1-x}$N layer.

To increase the carrier density, samples were illuminated through the optical access window of the magneto-cryostat for short periods with a flashlight and $\rho_{xx}$ monitored. After each illumination SdH and Hall measurements were performed with the sample kept in the dark. Typical SdH traces from sample I are shown in Figure 2(a). Under illumination, a significant increase in carrier density was observed: from $n=3.0\times10^{12}$ cm$^{-2}$ before illumination to $n=6.7\times10^{12}$ cm$^{-2}$ after the illumination had saturated the sample. The PPC effect in sample II was even more pronounced: carrier density changed from $1.6\times10^{12}$ cm$^{-2}$ to $5.9\times10^{12}$ cm$^{-2}$ under illumination. The corresponding electron mobility values were extracted from SdH and Hall data, which are plotted in Figure 2(b). As can be seen from this data, the mobility increases with increasing carrier density for both samples. At low carrier densities the dominant scattering is due to charged impurities and the scattering rate from charged impurities decreases with increasing Fermi wavevector. Furthermore, there is an enhanced screening with increasing carrier density [29]. The mobility-carrier density dependence observed in our samples is in agreement with the previous reports in AlGaAs/GaAs and AlGaN/GaN 2DEG structures [13, 26]. The enhancement in mobility with increasing carrier density is less significant at carrier densities higher than $4\times10^{12}$ cm$^{-2}$. Further increase in carrier density resulted in a saturation-like behavior near the end data points where maximum mobilities of 17830 cm$^2$/Vs and 21400 cm$^2$/Vs were recorded for sample I and sample II, respectively. No decrease in mobility was observed within the carrier density range of the experiment. We should note that in previous measurements performed in a wider carrier density range, it is found that the mobility first increased with carrier density, reached a maximum and then started to decrease [30]. The mobility reduction at higher carrier densities was mainly attributed to alloy disorder scattering and to interface roughness scattering at even higher (>7–8$\times10^{12}$ cm$^{-2}$) densities. Our non-decreasing high-mobility results confirm the effectiveness of the thin AlN interfacial layer: not only did it enhance the confinement of the 2DEG in the GaN channel, but it suppressed the alloy disorder scattering significantly as well. If we were able to increase the carrier density beyond $8\times10^{12}$ cm$^{-2}$, we would observe the drop in mobility due to interface roughness scattering and other scattering mechanisms effective in the high carrier concentration regime. To confirm the effectiveness of the AlN interfacial layer used in our samples, we have also grown and fabricated conventional Al$_{0.25}$Ga$_{0.75}$N/GaN heterostructures without an AlN layer. More than a two-fold improvement in mobility performance was achieved with the insertion of the AlN interfacial layer.

## B. Persistent photoconductivity measurements

The magneto-transport measurements demonstrated the presence of the PPC effect in our Al$_x$Ga$_{1-x}$N/AlN/GaN samples. With the intention of investigating this PPC effect, we conducted spectral illumination and thermal annealing experiments. Before illumination, in dark and at low temperature, the sample was in an equilibrium state with low carrier density and high sample resistance. With illumination, the sample state was changed to a non-equilibrium state, where the carrier density is higher and sample resistance is lower. The sample was then returned back to its original (or close to original) stable state by thermal annealing at sufficiently high temperatures. The illustration in Figure 3(a) shows the cycle used for the PPC experiments.

The conductivity evolution under the influence of illumination and annealing can be successfully described by defect configuration diagrams. Extensive PPC studies in the AlGaAs/GaAs system have been performed within this framework. Our purpose is to study and analyze the PPC effect in AlGaN/AlN/GaN heterostructures within the same framework with the goal of finding the threshold optical excitation energy for PPC as well as the critical annealing temperature needed for total recovery. Figure 3(b) shows a representative configuration coordinate diagram for a deep-level defect in $Al_xGa_{1-x}N$. Process #1 describes the optical excitation of the defect centers with incident photon energy higher than the threshold energy, $E_t$. The reverse process of annealing results in thermal capture of electrons back to the defect level, and is shown as process #2, where the critical thermal barrier to electron capture is $E_c$. It is possible that there might be a wide range of defects with different energy scales that participate in the PPC effect. To understand the energy scales involved in processes #1 and #2, we study the wavelength dependence of the PPC effect and perform annealing experiments at different temperatures, respectively. The results of illumination and annealing experiments are analyzed separately in the following sections.

## i. PPC - illumination experiment

In order to explore the spectral illumination dependence of PPC in our samples, we have designed the measurement setup shown in Figure 4. Single-wavelength illumination was achieved by using a 150W Xe-lamp, a long-pass filter, and a 30-cm monochromator. The sample was illuminated through the sapphire optical access window of our cryostat. The incident optical power was monitored using a calibrated UV-enhanced Si-detector (power meter). The sample excitation current was kept low at 1 μA to prevent heating. The longitudinal sample voltage and Hall voltage were measured using a low-noise pre-amplifier and lock-in amplifier. From these data, longitudinal resistivity ($\rho_{xx}$) and Hall resistivity ($\rho_{xy}$) values were determined. The samples were illuminated for 20 minutes at six different sub-bandgap wavelengths: 380 nm, 420 nm, 490 nm, 530 nm, 720 nm, and 870 nm. All measurements were done at $T = 4.5$ K. To block the possible higher harmonics, long-pass optical filters were utilized before the monochromator. Samples were annealed above room temperature after each illumination sequence to recover the equilibrium state.

In order to analyze the temporal PPC data, we have converted the resistivity curves into corresponding carrier density curves using the mobility vs. carrier concentration data obtained during magneto-transport measurements. Note that, in the context of the PPC effect, the carrier density is a more physically meaningful parameter than resistivity, as the change in carrier density is directly related to the number of defects that have changed by optical illumination. Figure 5 shows the typical change of carrier density with illumination at 420 nm, 490 nm, and 530 nm for both samples. Both samples exhibit an initial fast increase followed by a slower but steady enhancement. The rate of carrier density enhancement depends strongly on the illumination wavelength. The rate of change was over an order of magnitude when the wavelength was changed from 420 nm to 530 nm. Figure 5 inset presents the corresponding temporal change of resistivity of sample I under 420 nm illumination. Both samples showed decreasing PPC response as the illumination wavelength was increased. The PPC effect becomes very small at wavelengths longer than 530 nm.

To compare the PPC response curves at different wavelengths correctly, we need to normalize the rate of conductivity change with the incident photon flux. Therefore, we defined a dimensionless parameter, PPC-efficiency ($\eta$) as follows: the rate of electrons excited and contributing to the persistent photocurrent divided by the incident photon flux:

$$\eta = \frac{dn/dt}{I_o/h\upsilon},$$

where $dn/dt$ was obtained by differentiating the temporal carrier density curves in Figure 5, $I_o$ is the optical intensity incident on the sample surface, and $h\upsilon$ corresponds to the energy of a single incident photon. After taking the surface reflections into account, PPC-efficiency values were calculated and plotted as a function of photon energy (Figure 6). The efficiency dropped exponentially from 3.2 eV to 2.3 eV, for both samples at a similar rate. However, at 1.7 eV (720 nm) and 1.4 eV (870 nm) this trend was interrupted. A more complex PPC response was observed at these infrared wavelengths. An initial decrease in carrier density was noticed immediately after illumination started. Shortly after this negative photoconductivity response the carrier density started to increase slowly, indicating a positive PPC. The overall PPC response after 20 min was positive. At this point we are not sure about the origin of the initial negative photoconductive response under infrared illumination.

The maximum PPC-efficiency values achieved for sample I and sample II were 0.011% and 0.005% respectively. This means that 10,000 incident 3.2 eV-photons were able to excite at most one electron which contributed to the 2DEG channel persistently. To the best of our knowledge, this is the first time absolute quantum efficiency of PPC has been reported in this material system. Although this efficiency seems to be low, within 20 minutes, it changed the carrier density of sample I from $3.0 \times 10^{12}$ cm$^{-2}$ to $5.0 \times 10^{12}$ cm$^{-2}$. The inset of Figure 6 shows the PPC-efficiency curve of sample I in linear scale, where the significant drop is seen easily. For photon energies smaller than 2.3 eV, the pronounced PPC effect becomes very weak. We estimate $E_t = 2.0 \pm 0.2$ eV as the threshold excitation energy for PPC in Al$_x$Ga$_{1-x}$N/AlN/GaN heterostructures with x ≤ 0.25.

## ii. PPC - annealing experiment

The thermal annealing dependence of the PPC effect was studied in the same cryostat, where the sample temperature was changed with a controllable heater. Samples that were placed in a non-equilibrium state by optical illumination were annealed for 20 min, at temperatures ranging from 20 K to 280 K. After annealing, the temperature was ramped down to the initial temperature of 4.5 K, at which the final resistivity was recorded. The sample was then illuminated to set the sample in the same non-equilibrium state with same resistivity and carrier density before the next annealing step. The applied temperature profile for $T$ = 204.1 K and the resulting temporal variation of sample resistivity for sample I are shown in Figures 7(a) and 7(b), respectively. The initial reference resistivity for sample I before annealing was chosen as $\rho_{xx} = 57.2$ Ω which corresponds to $n_i \approx 6.2 \times 10^{12}$ cm$^{-2}$. After 20 min annealing, the final sample resistance was recorded as 60.5 Ω ($n_f \approx 5.9 \times 10^{12}$ cm$^{-2}$). The change in carrier density is found to be larger for higher temperature annealing.

We have also performed similar annealing experiments on sample II for which we started from $\rho_{xx} \approx 49.5$ Ω ($n_i \approx 5.9 \times 10^{12}$ cm$^{-2}$). After each annealing step, the sample resistivity was decreased to this value with illumination at 4.5 K. Figure 8(a) illustrates how the carrier density changed with annealing temperature for sample II. The sample did not recover completely even with annealing near room temperature, at 280 K. Either the annealing time or annealing temperature should be increased for complete recovery, i.e., to reach $n_f \approx 1.6 \times 10^{12}$ cm$^{-2}$. However, when compared with sample I, one can conclude that PPC in sample II is much more sensitive to annealing. Figure 8(b) shows the amount of carrier density reduction recorded as a function of annealing temperature. The change in carrier density for the maximum annealing temperature of 280 K is $0.9 \times 10^{12}$ cm$^{-2}$ and $3.2 \times 10^{12}$ cm$^{-2}$ for sample I and sample II, respectively.

Despite the differences in the amount of recovery we get at given annealing temperature, for both samples annealing happens for a broad range of temperatures. Thus, we could extract a single energy barrier for thermal electron capture from our data. It is likely that a range of defects with different energy barriers participate in the PPC effect.

At this point we do not have a definitive explanation for the microscopic mechanism responsible for the different PPC responses measured for Al$_{0.25}$Ga$_{0.75}$N/AlN/GaN and Al$_{0.15}$Ga$_{0.85}$N/AlN/GaN samples under the influence of illumination and annealing. However, we can merely suggest the following interpretation. Based on the previous reports on PPC effect in $Al_xGa_{1-x}As/GaAs$ and $Al_xGa_{1-x}N/GaN$, the origin of the PPC observed in these $Al_xGa_{1-x}N/AlN/GaN$ heterostructures may also be attributed to the excitation of deep level donors in the $Al_xGa_{1-x}N/AlN$ barrier layer. These deep levels can be compared with the DX centers in $Al_xGa_{1-x}As$ layers [12]. Since the AlN interfacial layer is very thin, it is suggested that deep level impurities in the 25-nm $Al_xGa_{1-x}N$ layer should be responsible for the PPC response. The difference in the intensity of PPC response for different Al-concentration supports this proposition. Furthermore, we have observed that the sample with lower Al-concentration, sample II, displayed a weaker PPC under illumination and a faster recovery under annealing when compared with sample I. This result suggests that the energy scales associated with the defect configuration coordinate diagrams should be different for these samples. For sample II, the stronger annealing response suggests that the thermal barrier to electron capture, $E_c$ is smaller. Since the threshold energy values are similar, the defect configuration coordinates in $Al_{0.15}Ga_{0.85}N$ and $Al_{0.25}Ga_{0.75}N$ defects should be very close. Additional experiments are needed to clarify the detailed PPC mechanism and to model the exact defect configuration coordinates in the corresponding $Al_xGa_{1-x}N$ layers.

## IV. SUMMARY

We have studied the spectral illumination and thermal annealing dependence of the persistent photoconductivity (PPC) effect in MOVPE-grown $Al_xGa_{1-x}N/AlN/GaN$ heterostructures. The PPC effect was used effectively to vary the carrier density and mobility in the 2DEG channel. SdH and Hall-effect measurements were performed to study the carrier density dependence of 2DEG mobility. The mobility increased with illumination and no decrease was observed till the PPC was saturated. A maximum mobility of $\mu = 21400$ cm$^2$/Vs at $T = 1.6$ K ($n=5.9\times10^{12}$ cm$^{-2}$) was achieved with the Al$_{0.15}$Ga$_{0.85}$N/AlN/GaN sample. The persistent photocurrent in both samples exhibited a strong dependence on illumination wavelength with threshold excitation energy of 2.0±0.2 eV. The PPC-efficiency of Al$_{0.25}$Ga$_{0.75}$N/AlN/GaN sample was higher than the efficiency of the Al$_{0.15}$Ga$_{0.85}$N/AlN/GaN sample. Carrier density in both samples increased with annealing, being stronger in the Al$_{0.15}$Ga$_{0.85}$N/AlN/GaN sample. However, a full recovery was not observed even at 280 K, showing that the PPC in these samples is effective even at room temperature. Comparing the measured results of two samples, the Al$_{0.25}$Ga$_{0.75}$N/AlN/GaN sample had a larger response to illumination, whereas it displayed a smaller change (recovery) with thermal annealing. These results suggest that the defect configuration-coordinate models for Al$_x$Ga$_{1-x}$N/AlN/GaN heterostructures depend strongly on the Al-composition of Al$_x$Ga$_{1-x}$N barrier layer.

## ACKNOWLEDGMENTS

This work was supported by the Air Force Office of Scientific Research under the direction of Dr. K. Reinhardt. The laboratory benefited from grants from the Office of Naval Research and the National Science Foundation and equipment funds provided by Virginia Commonwealth University.

# FIGURE CAPTIONS

**Figure 1.** (a) Longitudinal resistivity ($\rho_{xx}$) and Hall resistivity ($\rho_{xy}$) curves of $Al_{0.25}Ga_{0.75}N/AlN/GaN$ sample obtained from SdH and Hall measurements at $T = 1.6$ K. (b) SdH measurement results for sample I and sample II.

**Figure 2.** (a) Magnetoresistivity of sample I as a function of illumination. (b) Carrier density dependence of 2DEG mobility for both samples under study. All data were measured at $T = 1.6$ K.

**Figure 3. (a)** Schematic illustration of the PPC-illumination and annealing experiment and how the sample state changes during these measurements. (b) Configuration-coordinate diagram of a deep-level defect responsible for the PPC effect.

**Figure 4.** Schematic diagram of the measurement setup used for PPC experiments.

**Figure 5.** Temporal carrier density curves of sample I and sample II for different illumination wavelengths. Inset shows the temporal evolution of resistivity for sample I at illumination wavelength of 420 nm. The measurements were performed at 4.5 K.

**Figure 6.** PPC-efficiency of the $Al_xGa_{1-x}N/AlN/GaN$ heterostructures as a function of incident photon energy. Inset shows the efficiency curve of sample I on a linear scale.

**Figure 7.** (a) Temperature profile for annealing measurement at $T = 204.1$ K. (b) The resulting temporal variation of the resistivity for sample I.

**Figure 8.** (a) The carrier density of sample II. (b) Comparison of the change in carrier density as a function of annealing temperature for sample I and sample II.

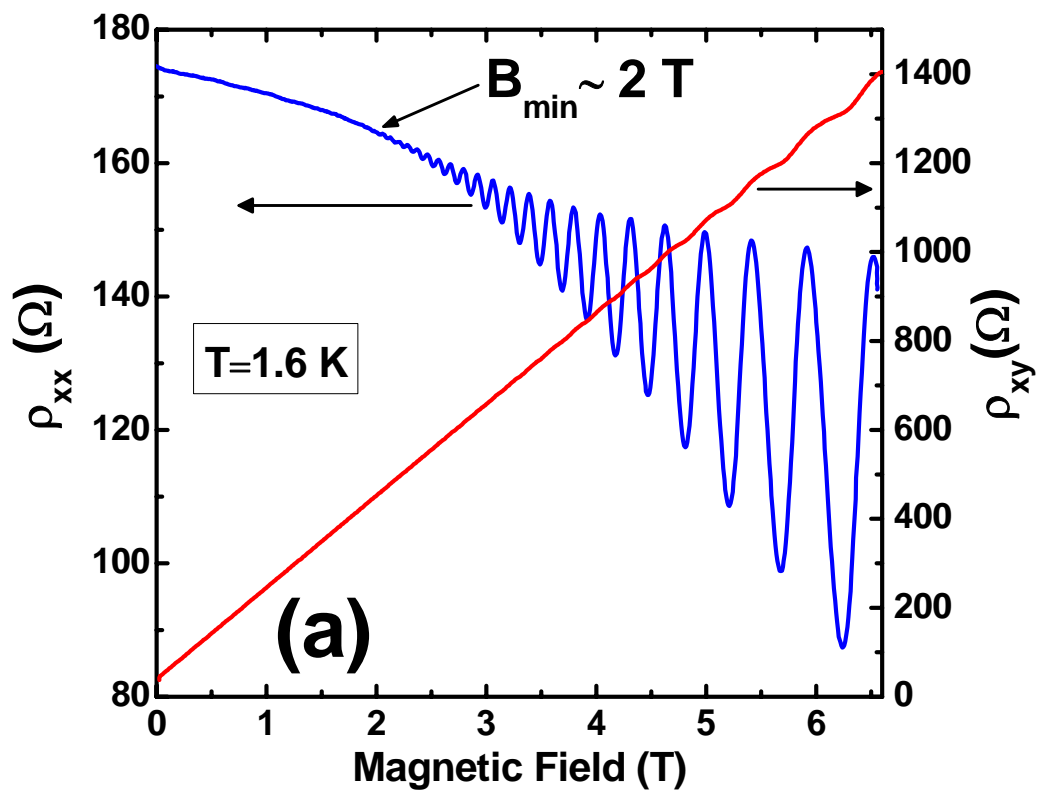
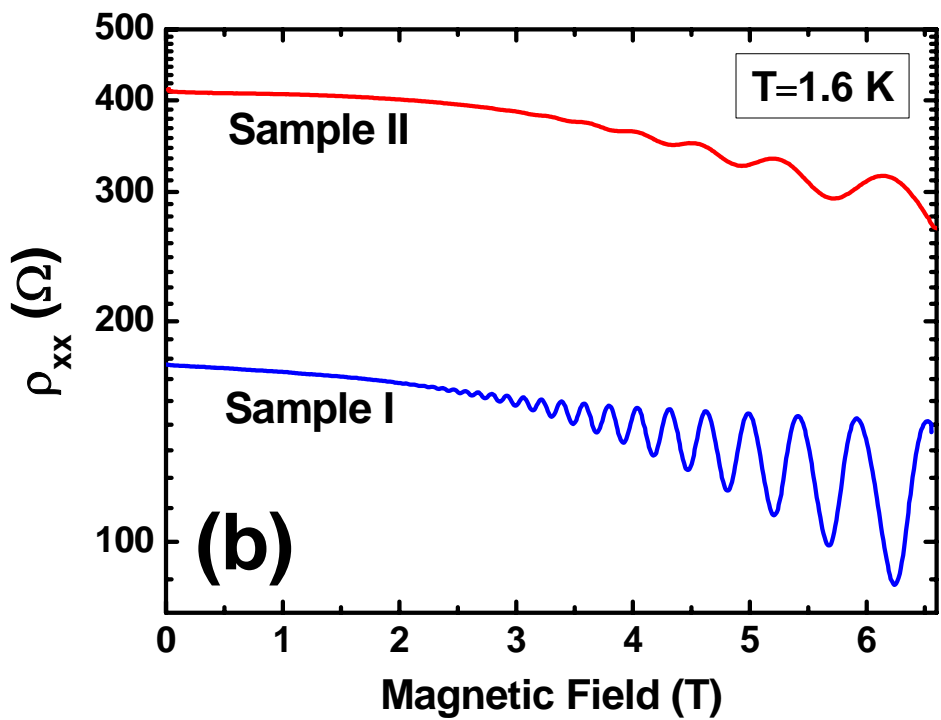

Biyikli *et al.* **Figure 1**

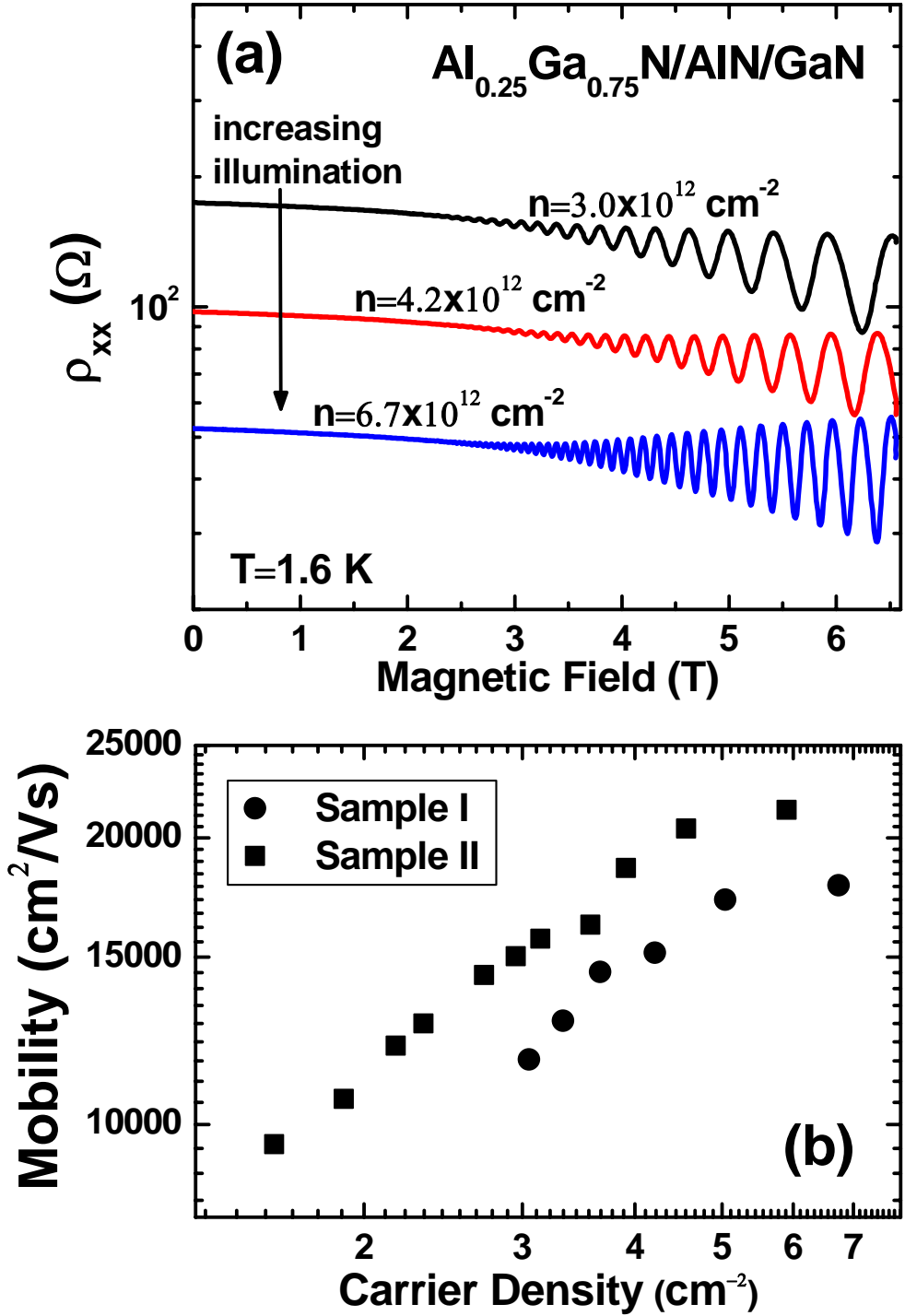

Biyikli *et al.* Figure 2

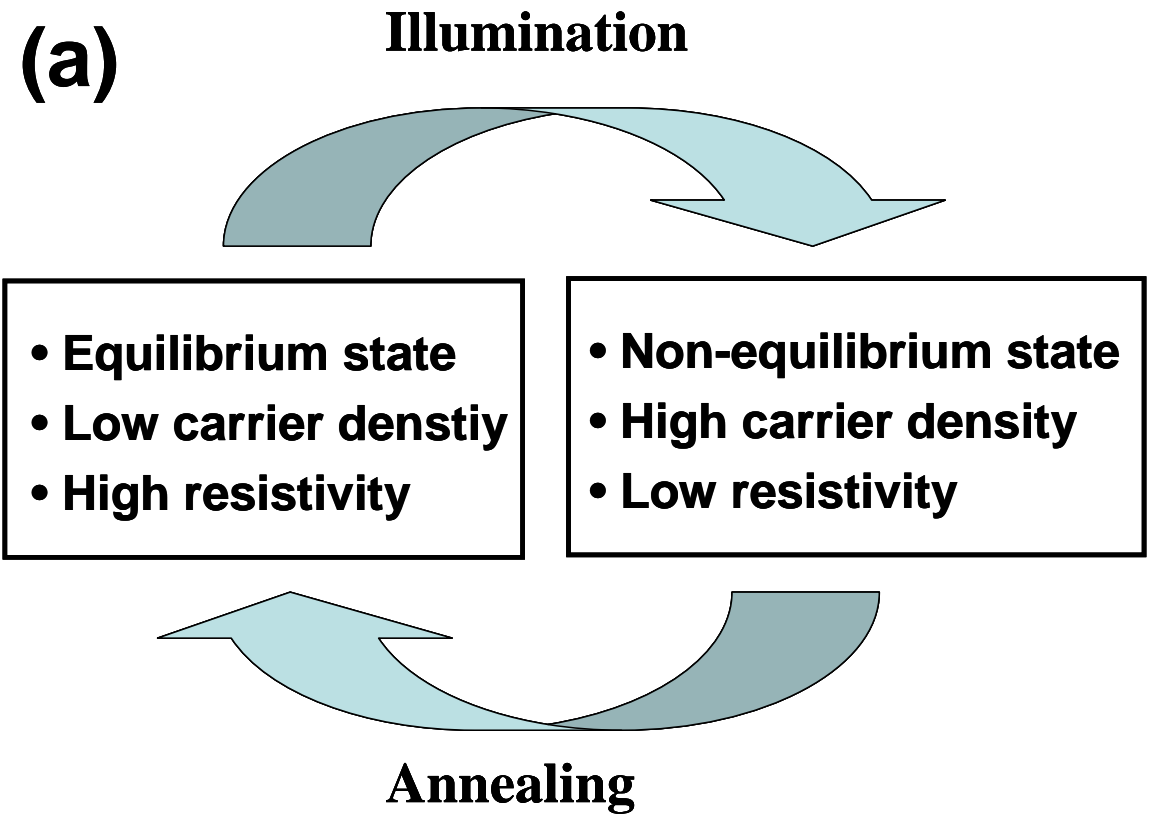
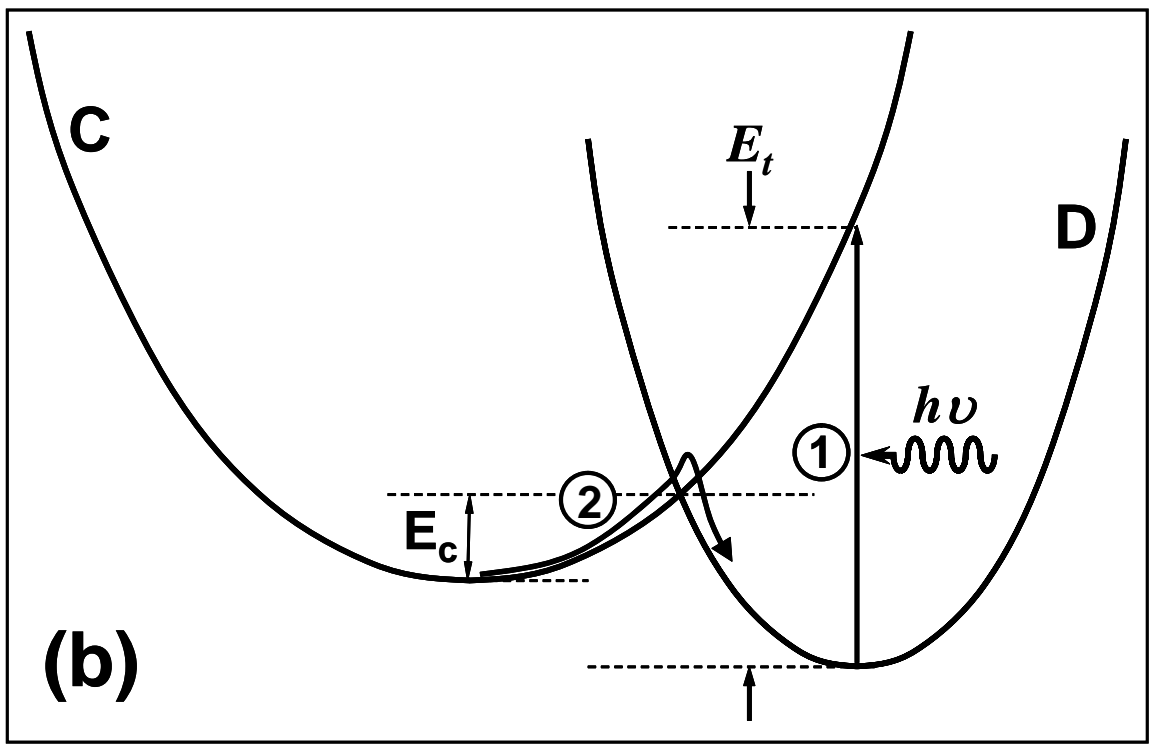

Biyikli *et al.* **Figure 3**

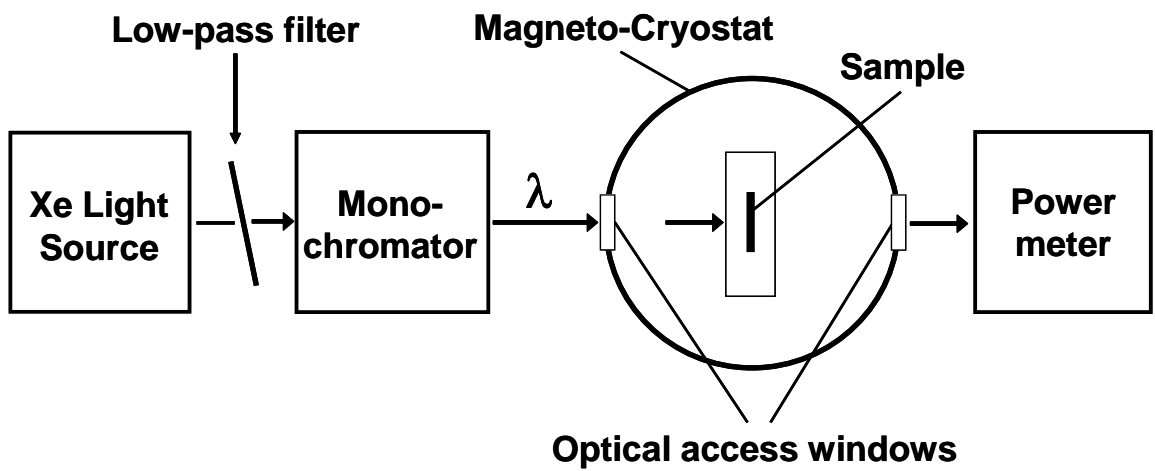

Biyikli *et al.* **Figure 4**

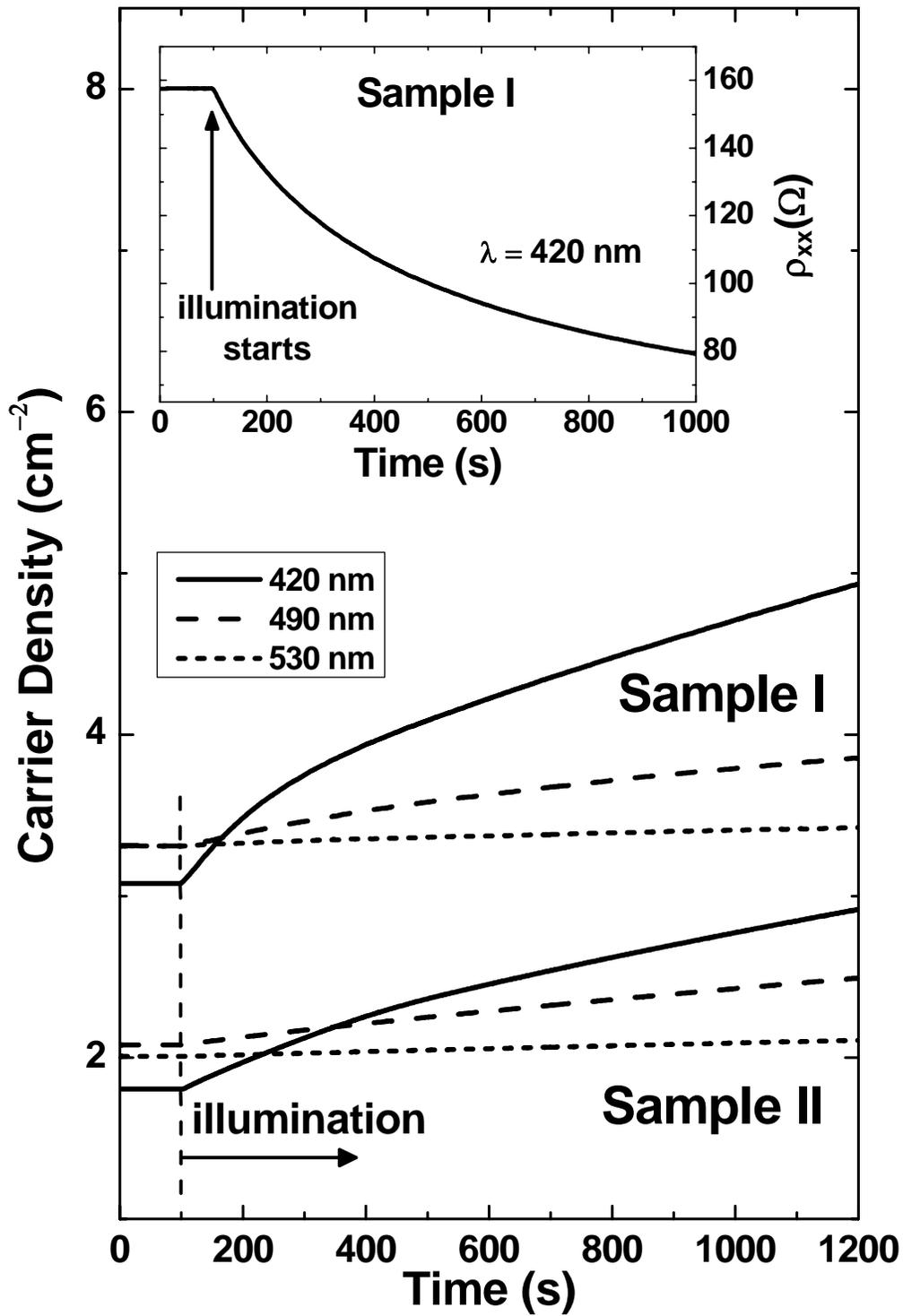

Biyikli *et al.* **Figure 5**

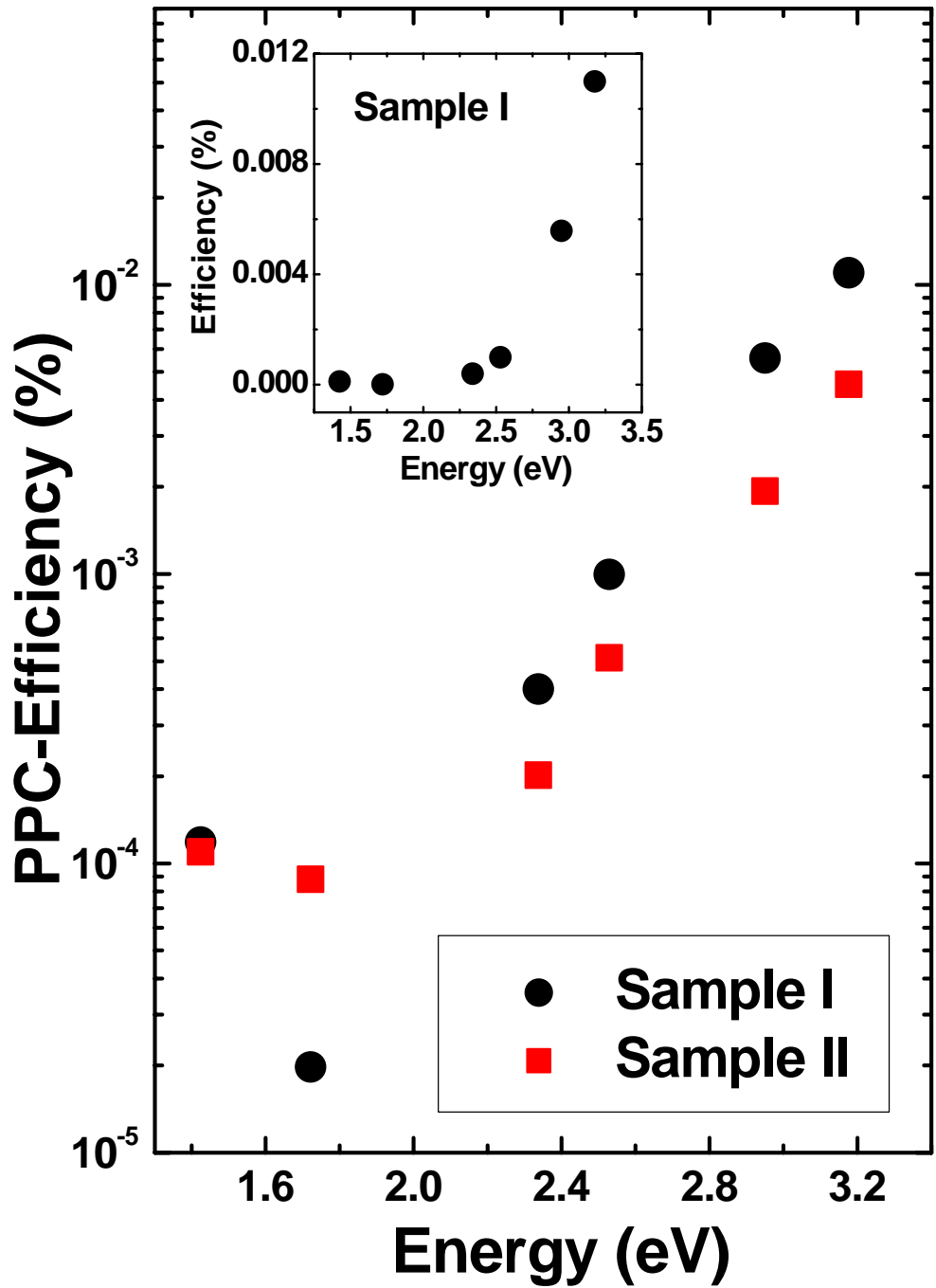

Biyikli *et al.* **Figure 6**

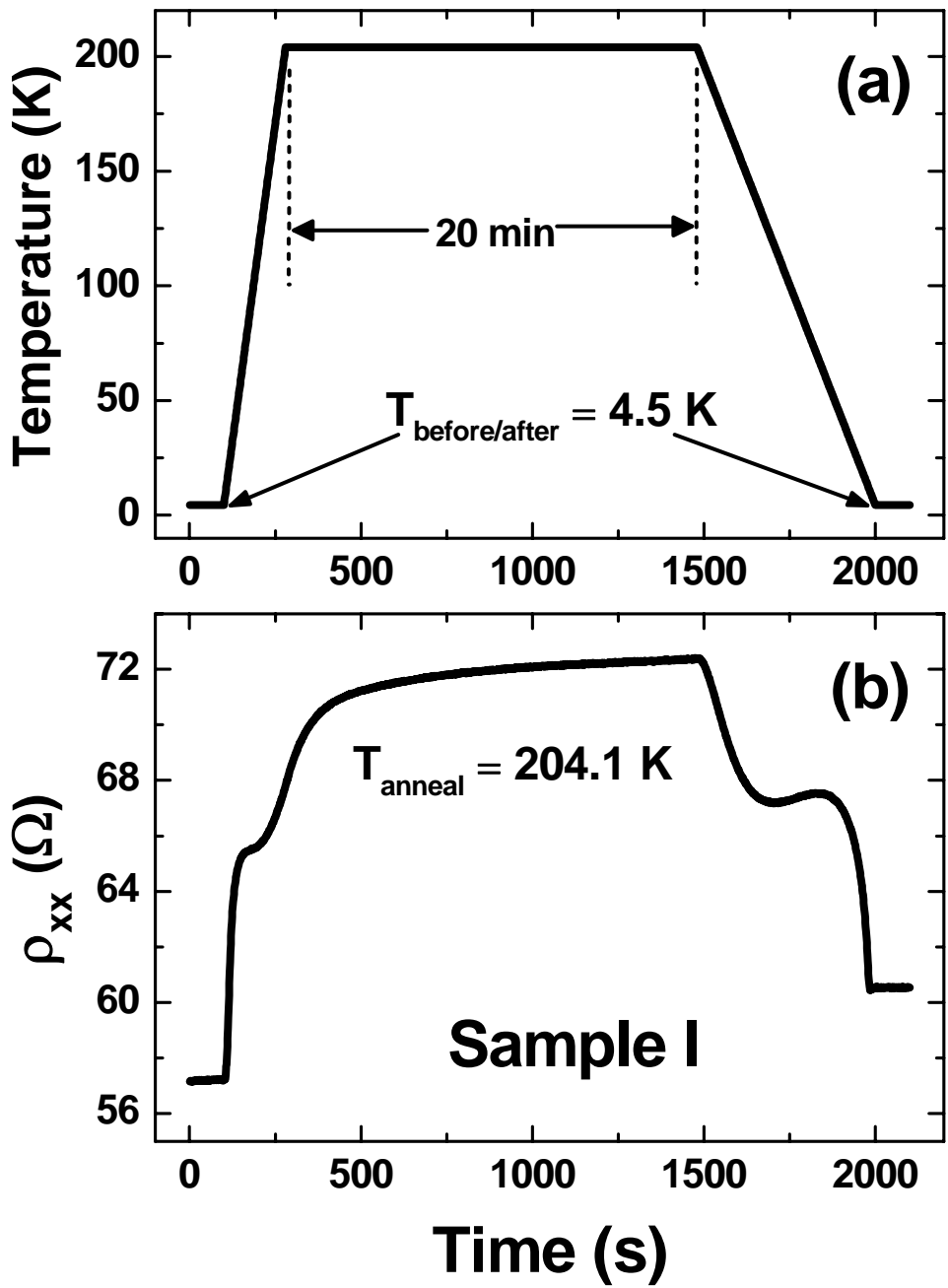

Biyikli *et al.* **Figure 7**

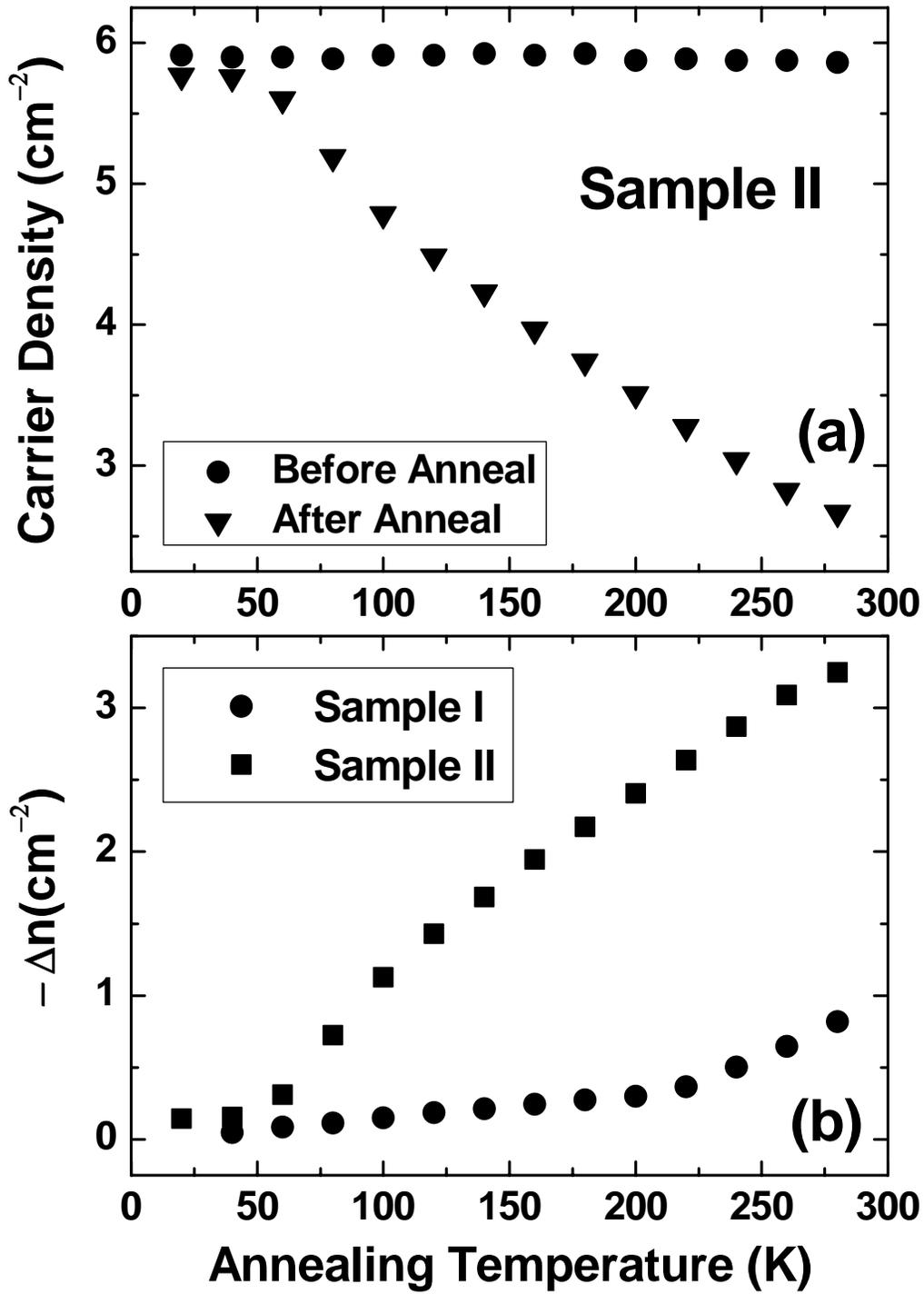

Biyikli *et al.* **Figure 8**